# Bimodal star formation in elliptical galaxies and the enrichment of the intra-cluster medium

D. Elbaz[1], M. Arnaud[1], E. Vangioni-Flam[2]

[1] C.E.A., DSM, DAPNIA,Service d'Astrophysique, C.E. Saclay, F-91191, Gif-Sur-Yvette Cedex, France
[2] Institut d'Astrophysique de Paris, CNRS, 98 bis Bd Arago, 75014 Paris, France

**Abstract.** A detailed model of galactic evolution is proposed to explain both the iron content of the Intra-Cluster Medium (ICM) and of the ellipticals pertaining to the cluster. SNII are responsible for the production of Fe in the galaxies and of its partial ejection in the surrounding medium. A high formation rate of massive stars only ($m > 3\ M_\odot$) at the very early stage of evolution of ellipticals, reminiscent of starburst galaxies, is advocated. The high rate of SNII explosions induces a powerful galactic wind, followed by a more quiescent evolution where stars are formed, with a normal Initial Mass Function (IMF), from the enriched gas remaining after the wind phase and from the gas restored by the longest living stars produced during the early phase. Most of the iron is produced by the high-mass stars created in the early burst phase.

The ICM iron mass per unit luminosity of member galaxies ($2 \times 10^{-2}\ M_\odot/L_\odot$) is reproduced, together with a mean $Mg/Fe$ ratio in the stellar component higher than solar, and other observed quantities on present elliptical galaxies ($M/L$ ratio, stellar and ISM metallicities, remaining gas fraction). We discuss the influence of the parameters: astration rate in the burst phase and in the quiescent phase, mass loss rate during the wind phase. We found that the global iron production is very robust with respect to the parameters. On the other hand, a key factor is the total mass lost during the wind. It must be $\approx 50\%$ in order to insure a correct share between present stars and the ICM. The evolution depends on the galaxy mass due to the deeper gravitational potential for larger galaxies. A possible weak point of the model is that, although metallicities increase with galaxy mass, other quantities ($Mg/Fe$ and $M/L$) vary only slightly.

It is concluded, as in previous works, that most of the cluster gas (80 % for rich clusters) is of primordial origin, which explains the decrease of iron abundance with cluster temperature, itself related to the richness of the cluster. We also predict that the abundances of O and Si in the ICM, relative to Fe are higher than solar.

Send offprint requests to: D. Elbaz

## 1. Introduction

The hot X-ray emitting intra-cluster medium (ICM) is enriched in heavy elements: the iron abundance is about 0.35 solar (Rothenflug & Arnaud 1985; Edge and Stewart 1991) and decreases slightly with the ICM temperature (Hatsukade 1989; Arnaud et al. 1992; Ikebe et al. 1992; Tsuru 1993). Part of the ICM must have been processed in the stars of the galaxies and ejected into the intergalactic space to account for this observation. The difficulty but also the interest in modeling the enrichment of the ICM comes from the fact that it involves a large variety of important astrophysical issues, as emphasized in the recent paper of Renzini et al. (1993), at all astrophysical scales: stellar evolution and supernovae physics (the SN being the only iron producers), star formation, formation and chemical/dynamical evolution of galaxies and environmental effects, origin of the ICM and cluster formation. If a general consensus has emerged that the ICM iron originated from elliptical galaxies and was ejected through SN driven wind/outflows, the relative role of type I and type II SN is still a matter of controversy (Matteucci & Tornambé 1987; Matteucci & Vettolani 1988; Matteucci 1992; David et al. 1990b, 1991a,b; Ciotti et al. 1991; White 1991; Arnaud et al. 1992; Hattori & Terasawa 1993; Renzini et al. 1993).

In this paper, we propose a model for the evolution of elliptical galaxies, based on bimodal star formation as suggested by Arnaud et al. (1992) to explain the huge amount of iron present in clusters. In this model, most of the heavy elements are produced by SNII during a first violent phase where only high-mass stars are formed (high-mass mode). This early phase ends with a strong galactic wind driven by SNII when the specific thermal energy exceeds the gravitational binding energy of the interstellar gas. During this wind phase we assume that star formation stops, due to a too hot environment. This phase will typically last 0.04 Gyr, the lifetime of an $8M_\odot$ star: when the last SNII has exploded, the specific thermal energy decreases rapidly due to both cooling and mass input from longer lived stars. A new star formation phase will then follow, where stars are formed from the pre-enriched gas left over from the wind phase and from the gas restored by the long lived stars produced during the high-mass mode. In this phase, we assume a more quiescent star formation, with a normal IMF (normal mode, with a Salpeter IMF and a star formation rate proportional to the gas mass fraction). There are thus three main parameters in our model: the star formation rate in each phase and the mass loss rate during the wind phase (our model does not include a detailed hydrodynamical modeling of the wind). In section 2, we discuss qualitatively the advantages of this model in the general framework of the ICM enrichment and the evolution of galaxies in general. The observational constraints (for both the galaxies and the ICM) to which we will confront this type of model are summarized in section 3. In section 4, the basic physical ingredients of our model and the method used are described in detail. In section 5.1, we present and discuss, in a quantitative way, the dynamically and chemical evolution of an elliptical galaxy of $5 \times 10^{11} M_\odot$ for a "standard" set of parameters, defined in an empirical way. How the final results, in terms of each observable quantity, are controlled by the parameters is discussed in section 5.2. We also will see that good agreement with observations is obtained for plausible values of the parameters; such agreement does not just stem from the fact that we used a complex multi-parameter model as compared to the available constraints, which is always a risk in modeling. In section 6, we consider the dependence of the results on the galaxy mass. In section 7, we integrate the gas and iron mass ejected into the ICM by the galaxies and discuss the origin of the ICM.

## 2. The problem of the heavy element enrichment in clusters

Reliable constraints can now be put on the enrichment process, due to precise estimates, via X-ray imaging (gas mass determination) and spectroscopy (iron abundance determination), of the iron mass present in the ICM. It must be recalled that the iron abundance by itself is not a meaningful constraint: it is a quantity that depends on the relative proportion of primordial ICM gas, in which the iron mass ejected by the galaxies is diluted. A model of the ICM enrichment, consistent with the iron mass present in the ICM, will tell us what part of the ICM comes from the galaxies and what part is primordial, a fundamental quantity for understanding the formation of galaxies and clusters.

To provide a real constraint, the iron mass present in the ICM must be compared (or normalized) to a global quantity, characteristic of the stellar component of the cluster, from which the iron is produced. Arnaud et al. (1992) showed that the iron mass contained in the ICM is directly proportional to the luminosity of elliptical and S0 galaxies. This correlation was recently confirmed by the latest results of the Ginga satellite (Arnaud 1994, Arnaud et al. 1994). This simple proportionality law naturally led to the conclusion that the ICM iron originates in E/S0 galaxies and that it is ejected through galactic winds. Other ejection processes, like ram pressure stripping, which depends on the environment, would imply a general variation of the iron mass over luminosity ratio with cluster richness (Renzini et al. 1993). This conclusion is not surprising in view of our current understanding of ellipticals. Ellipticals are well known to contain an old stellar population and a low proportion of gas. These two observations are usually explained by considering that ellipticals have experienced a violent burst of star formation in the early stage of their evolution. The strong concentration of supernovae expected at the beginning of this evolution is likely to have driven an outflow, resulting in the metal enrichment of the ICM.

If the general framework for the enrichment process seems reasonably understood (SN driven outflow from elliptical galaxies), problems arise when one considers quantitatively the mass of iron present in clusters (both in the ICM and in the stellar component of the galaxies). Arnaud et al. (1992) showed that the iron mass over stellar mass ratio is about 4 times larger in clusters than in our Galaxy. If a constant Initial Mass Function (IMF) is assumed, the iron mass to stellar mass ratio is directly linked to a fundamental quantity in chemical models: the iron mass that a given population of stars can produce or the iron global yield, which depends only on the IMF. In that case one cannot account for the iron present in clusters by a "standard" iron yield, i.e. one adapted to our Galaxy. This is a real challenge for any enrichment model. It must be noticed that the estimate of the iron mass over stellar mass ratio from the iron mass over luminosity ratio, which is the quantity directly observed, requires a knowledge of the mass-to-light ratio, a quantity somewhat uncertain (see section 3). Considering directly the total iron mass to light ratio, Renzini et al. (1993, see also Ciotti et al. 1991) actually reached a similar conclusion: a high past supernova activity is required (i.e. high iron global yield), either a much higher SNIa past average rate as compared to the present value observed in ellipticals (at least a factor of 10) or a high SNII rate (due for instance to a very flat IMF). Thus the main issue at hand is to understand the relative importance of SNIa and SNII in the enrichment process (for both the gaseous and stellar components of clusters) and why their integrated rate has been so high. This is specially difficult because fundamental information on the SN themselves are lack-ing. If we know with some confidence how much iron is produced by SNIa, their rate in the past is uncertain due to the physics of their progenitors (white dwarfs in binary systems): it thus depends on the binarity and separation of the two stars. Conversely for SNII, their progenitors are known (stars more massive than 8 $M_\odot$) and their rate is a direct consequence of the IMF and star formation rate assumed but their Fe yield is uncertain.

We think that there are converging arguments, although none of them are fully conclusive by themselves, that SNII play the major role both in the ICM enrichment and the establishment of the stellar metallicity observed in present-day ellipticals. From an observational point of view, the high $(O/Fe)$ ratio observed in the ICM of Virgo and Perseus (Canizares et al. 1982, 1988) already pointed toward a metal enrichment mainly by SNII. Indeed $O$ is only produced by SNII while $Fe$ can be produced by both SNIa and SNII. Ratios above solar are typical of enrichment by SNII since SNIs play the major role for the iron enrichment in our Galaxy. This relative overabundance of Oxygen seems now confirmed by the last ASCA observations (Mushotzky 1994). Moreover X-ray observations of the hot corona of the bright elliptical galaxies NGC 4472, NGC 1399 and NGC 4636 indicate iron abundances less than twice solar, with best fit values around solar (Ikebe et al. 1992; Serletmitsos et al. 1993; Forman et al. 1993). Forman et al. (1993) argue that these results are unconsistent with models assuming high SNIa rates in the past which predict much higher iron abundances (see also Renzini et al. 1993 for alternative explanations). Recent ASCA data (Awaki et al. 1994) tend to confirm this low metallicity with even smaller best fit values: 1/2 solar or below. On the other hand, in order to account for the ICM Fe content, i.e. to insure both a correct global iron production and its further ejection, the models based on SNIa require a rapidly decreasing SNIa rate. Ciotti et al. (1991) have empirically parameterized the SNIa rates as $R_{SNI}(t) \alpha t^{-s}$ and concluded that $s > 1.4$. Other enrichment models are directly based on a specific modeling of the SNI progenitors (Greggio & Renzini 1983), where SNI are supposed to arise from C-deflagration in white dwarfs in binary systems and some assumption is made on the mass ratio distribution function of the binary. They predict a much weaker time dependence of the SNI rate (David et al. 1991b; Matteucci & Tornambé 1987). In that case it seems unlikely that the iron produced, although in enough quantity, could be actually ejected from the galaxy, due to less efficient heating. David et al. (1991b) showed that a high SNII rate (flat IMF) in the past is further needed to explain the high Fe content of the ICM (the iron produced by SNIa is not entirely ejected); Matteucci (1992) also found that the thermal energy due to SNIa, after the early wind driven mainly by SNII, is always smaller than the binding energy.

Finally, the stellar metallicity, although not precisely known (see next section), seems to be above solar, at least in massive ellipticals, with a $(Mg/Fe)$ ratio about twice solar. This overabundance is expected if SNII are a major contributor to the stellar metallicity, and again a high SNII rate seems to be required. If a normal IMF is assumed, detailed chemical evolution models predict stellar Fe abundances below solar (Matteucci & Tornambé 1987). A flatter IMF for more massive galaxies would be in better agreement with the variation of the $(Mg/Fe)$ ratio with galaxy mass (Matteucci 1993).

If SNII have indeed played the major role in the ICM enrichment, it would be interesting to understand why (rather than to simply assume an ad-hoc flat IMF). One major advantage of the standard picture, assumed by most authors until now (normal IMF and major contribution from SNIa, as in our Galaxy), was its simplicity; moreover it left open the possibility of a universal IMF, as emphasized by Renzini et al. (1993). Actually there are already indications, both in our Galaxy and in starburst galaxies, that the IMF may not be universal and, in particular, is dependent from the Star Formation Rate (SFR), through the physical conditions in the star formation regions. The observed characteristics of starburst galaxies seem to be fitted better by models where only high mass stars are produced with a lower cut-off for the IMF at $M \gtrsim 3 M\odot$ ( Wright 1988; Rieke et al. 1993; Doane & Mathews 1993; Charlot et al. 1993 and references therein). This is supported by some theoretical arguments: both the characteristic stellar mass and star formation efficiencies should increase with the internal turbulent motion of the formation region, with an additional feedback from high-mass stars due to energy input from the high SNII rate associated with a high SFR (Silk 1993, 1995 see also Henriksen 1991). SNII blast waves could also by themselves destroy any forming low mass stars (Doane & Mathews 1993). If this picture is valid, such a truncated IMF should also apply at the beginning of the evolution of elliptical galaxies (phase I of our model), supposed to have experienced an even more intense starburst phase at the scale of the whole galaxy (Silk 1993). This type of model, that we propose, has also the advantage a priori that a very high SNII rate is obtained in the early stage of evolution, producing both iron and thermal energy very rapidly and thus inducing a SN-driven wind of enriched gas while the gas mass fraction is high and little of the iron blocked into stars. We can thus expect to account for the ICM enrichment. However, since we observe long lived stars ($M \lesssim 1 M\odot$) in present-day ellipticals, this first phase must have been followed by a more quiescent phase, producing stars in the whole mass range (our phase III), requiring some gas to be still available after the wind phase. Star formation will thus proceed from a pre-enriched gas. If the galaxy loses about half of its mass during the wind and since most of the iron is produced during the early phase, due to the much higher iron yield there, we naturally expect roughly the same amount of iron in the stellar component as in the ICM of clusters. This is exactly what is observed (Ar-naud et al. 1992; Renzini et al. 1993); a second advantage of this model. The relationship between the ejected gas mass and the distribution of the iron mass between the galaxies and the ICM is not so natural in the case of a mixed contribution of SNII and SNIa because of their different timescales. Larson (1986) has originally proposed a bimodal star formation model for our Galaxy, which linearly combines a high-mass mode, which forms only high-mass stars and which is preponderant at early times with a normal mode which forms stars of all masses at a nearly constant rate. This model was proposed to account for the unseen mass that has been thought to exist in the solar neighbourhood. Now this argument is obsolete (Larson 1991). Nevertheless a bimodal star formation could be apply elsewhere provided the massive mode is adjusted in different astrophysical contexts. In particular François et al. (1990) have showed that if the two modes appear sequentially (first the high-mass mode with an exponentially decreasing rate, then the normal mode) all the main observational constraints on our Galaxy are satisfied, using a moderate high mass mode: the G-dwarf metallicity, the age-metallicity distribution and the abundances of Deuterium and other isotopes.

In conclusion bimodal models could be an alternative way to account for the observations of our Galaxy, starburst galaxies, elliptical galaxies as well as of the ICM. Before presenting our quantitative results, we shall briefly review in the next section the observational constraints on the ICM and elliptical galaxies that we will use afterwards to validate our model.

## 3. Observational constraints

### 3.1. Cluster scale

Using a sample of clusters observed by GINGA Arnaud et al. (1992, see also Arnaud 1994) showed that the iron mass present in the ICM, $M_{Fe}^{ICM}$ is correlated with the total luminosity of E/S0, $L_v^{E/S0}$. The best fit corresponds to a simple proportionality with:

$$\frac{M_{Fe}^{ICM}}{L_V^{E/S0}} = 2 \times 10^{-2} \frac{M_\odot}{L_\odot} \qquad (1)$$

There is some scatter in the correlation between these two quantities: $M_{Fe}^{ICM}$ can vary by a factor of two for a given $L_V^{E/S0}$. This deviation from the best fit law, which does not present any obvious trend with cluster luminosity is probably due to the uncertainties on the gas mass determination. The iron mass, the product of the iron mass fraction and the gas mass, has been computed using the very precise overall iron abundance determined by GINGA and the gas mass derived from EINSTEIN or ROSAT data.

The main uncertainties come from assuming a uniform iron abundance in the ICM and from the uncertainties on the actual extent of clusters. Global spectra as obtained by GINGA provide the emission weighted mean iron abundance. Were the iron distribution more peaked than the gas distribution we would then overestimate the total iron mass. Spatially resolved spectroscopy is still scarce, but rapidly increasing with the new results from the ASCA satellite. Before ASCA launch such data were available only on Coma (Hughes et al. 1993), Perseus (Ponman et al. 1990) and Virgo (Koyama et al. 1991). Whereas no iron gradient seems to exist in Coma, there were some indication of an increase of iron abundance at the very center of the two other clusters. However, as discussed by Hughes et al. (1993), the central emission of these two clusters is complex and dominated by the central galaxy and cooling flow emission, which is not the case for Coma. This could well lead to large uncertainties in the iron abundance determination. First results from ASCA (Mushotzky 1994; Ohashi et al. 1994; Ohashi 1994 and references therein) seem to confirm that the iron abundance is constant, in particular for Perseus (but not for Virgo). Till now a metal concentration has been observed only at the center of few poor clusters. Note that the origin of this central concentration is not well understood and that a specific enrichment by the central cD may have to be invoked (Ohashi 1994).

Concerning the second source of uncertainty, the ratio of gas mass over luminosity depends on the assumed extent of the cluster atmosphere (it increases with radius), since the galaxy distribution is more peaked than the gas distribution. A cut-off at about 10 core radii has been chosen (3 Mpc for rich clusters), which is, for most clusters, beyond the maximum radius where the X-ray emission is detected significantly. Thus the gas mass could have been overestimated. However recent ROSAT data on bright clusters confirm that the gas extends very far from the cluster center: at least up to 3-4 Mpc (Schwarz et al. 1992; Briel et al. 1992; Henry et al. 1993).

Finally, the ratio of the total ICM mass to $L_v^{E/S0}$ provides an upper limit for the ejected gas mass deduced from the model. In practice, this limit is never reached and this constraint is used to deduce the proportion of primordial gas present in the ICM. Arnaud et al. (1992) showed that this ratio increases from groups to rich clusters in the range:

$$20 \ \frac{M_\odot}{L_\odot} \leq \frac{M_{ICM}}{L_V^{E/S0}} \leq 50 \ \frac{M_\odot}{L_\odot} \qquad (2)$$

Similar results were obtained by David et al. (1990a) and Tsuru (1993) on the whole galaxy population.

### 3.2. Galaxy scale

#### 3.2.1. Metallicity of nearby ellipticals

Data on stellar metallicity are derived indirectly from Mg and Fe line strengths and some empirical or/and theoret-ical calibrations. Buzzoni et al. (1992) derived the following relationship between the $Mg_2$ index and the iron abundance: $Mg_2 = 0.135\,[Fe/H] + 0.28$, where $[Fe/H]$ is the usual logarithmic abundance relative to the sun. They derived a mean iron abundance in ellipticals of $[Fe/H] = 0.15$, corresponding to a $Mg_2$ index of 0.3, with a large spread of $\pm 0.5$ among the galaxy population.

It is well established that the $Mg_2$ index is tightly correlated with the central velocity dispersion (e.g. Bender et al. 1993). It indicates that more massive galaxies should have a higher metallicity. However the direct correlation between $Mg_2$ and luminosity shows a considerably larger scatter (Vader 1986) and we did not extrapolate the $[Fe/H]$ versus $L$ relationship from the $Mg_2$ versus $L$ and $[Fe/H]$ versus $Mg_2$ relationships. We simply considered that a galaxy of luminosity $L_V = 3 \times 10^{10}\,L_\odot$, which corresponds to the $Mg_2$ index = 0.3 (Vader 1986)[1] and has a luminosity close to the break luminosity in the Schechter luminosity distribution (see section 7), should have a stellar logarithmic iron abundance of $0.15 \pm 0.10$. The error bars given are empirical and estimated taking into account the typical scatter in the $Mg_2$ versus $\log(L)$ correlation. It also encompasses the variations linked to the exact choice of the calibration slope (see Worthey et al. 1992).

Further constraints on the variation of metallicity among galaxies are actually given by the recent data on $Mg_2$, $Fe5270$ and $Fe5335$ index (Worthey et al. 1992), covering a $Mg_2$ index range of $0.2 - 0.35$, i.e. more than 2 orders of magnitude in luminosity. They seem to indicate that Magnesium is overabundant with respect to Iron in giant ellipticals (with again a large scatter): $[Mg/Fe] = 0.2 - 0.3$ and that it increases with galaxy luminosity or mass: the variation of $[Fe/H]$ should be about half that of $[Mg/H]$.

The measurements on the metallicity in the hot gaseous component of ellipticals, made in X-ray, are scarce (see references in sections 2) and possibly in contradiction with measurements on stellar metallicities. Indeed, if the very low values (half solar) obtained from ASCA measurements are further confirmed, we shall be facing a serious problem: no chemical evolution model (with no inflow) can accommodate a metallicity in the ISM (below solar) lower than in the stellar component (above solar, see above). For galaxies inside clusters (but obviously not for isolated galaxies), this discrepancy could be indicative of an important dilution effect by accretion from the ICM (as discussed in Renzini et al. 1993). There might also be a serious problem in the stellar metallicity determination. For the moment, we simply note that the sample studied by ASCA is still small and only simple models for the galactic halos have been considered yet. In view of these uncertainties, we will adopt the most precise ROSAT measurements (previous to ASCA) of Forman et al. (1993) on NGC 4472: an iron abundance between one and twice solar.

### 3.2.2. Mass over luminosity ratio and gas content

The mass to light ratio ($M/L$) is an important constraint because it depends on the IMF. Enhanced formation of massive stars in more massive galaxies, leaving more remnants, has actually been invoked to explain the apparent increase of $M/L$ with mass (Larson 1986). The estimate of the luminous mass relies on the virial theorem and observed effective radius and velocity dispersion, assuming isotropic velocity dispersion, mass follows light, spherical galaxy following the $R^{1/4}$ law. It could be contaminated by the presence of dark matter in the core. For a galaxy of $L_V = 3 \times 10^{10} L_\odot$, we derived from Bender et al. (1992, Fig.1, sample of Virgo and Coma ellipticals) a typical $M/L_V$ ratio of 6, with individual values ranging from 3 to 9 for galaxies of that luminosity. The $1\sigma$ dispersion is about 20% for the whole sample and about 12% if one discards some peculiar ellipticals (Renzini and Ciotti 1993). Somewhat smaller values, $M/L_V \sim 3$, are obtained by Van der Marel (1991) who used more sophisticated dynamical models. The possible presence of a dark component has been studied by Saglia et al. (1992), but it appeared that the mass-to-light ratio of the luminous component is not changed dramatically $M/L_V \sim 5$ (because the amount of dark matter in the central part is small). Thus we will simply assume in the following that $M/L_V = 6 \pm 3$ for a $3 \times 10^{10} L_\odot$ elliptical galaxy.

The variation of $M/L$ with luminosity is uncertain. It seems to slightly increase with luminosity, $M/L_V \alpha L_V^{0.2-0.3}$ (e.g. Vader 1986; Bender et al. 1992). However there is a scatter in the $M/L$ versus $L$ correlation, which is still consistent with a constant $M/L$ ratio (Lauer 1985; Djorgovski & Davis 1987). Moreover as discussed by Bender et al. (1992) the existence of this variation relies on the assumptions made on the structural properties of the galaxies. They could vary with the galaxy mass so that actually $M/L$ may be constant.

Finally most of the gas in elliptical galaxies is in the form of a hot halo, as revealed by X-ray measurements. The gas content is low, with typical gas over stellar mass ranging from 0.01 to 0.07 (Forman et al. 1985).

## 4. The bimodal star formation model for the evolution of elliptical galaxies

### 4.1. The evolution model

The galaxies are supposed to be formed from a cloud of gas with no pre-enrichment and the star formation is assumed to occur on a short time scale, following the stan-dard picture of ellipticals, described in section 2. At that stage it may be useful to discuss further these first assumptions. The reason for the different evolutionary scenarios expected for spirals and ellipticals is not clear up to now. It has been proposed that ellipticals result from the merging of spirals or proto-spirals that would provoke violent star formation.

---

[1] The results on the ICM were obtained by Arnaud et al. (1992) assuming $H_0 = 50\,km/s/Mpc$ and $L_V \sim 1.2 L_B$. When necessary, as it is the case here, the results from other authors are scaled accordingly for consistency

If the merger involves already evolved spirals (with a star formation history) our model would not be valid anymore. De Carvalho and Djorgovski (1992) found that cluster ellipticals form a more homogeneous population than the field elliptical galaxies (less scatter in the Fundamental Plane correlations) which could be indicative that cluster galaxies form early from many gas-rich fragments, whereas at least some field galaxies form from more evolved system. If the merging objects in clusters have indeed nearly no previous star formation and thus nearly no pre enrichment, our model is still pertinent. The 't=0' moment of the bimodal star formation model presented here should then be understood as the beginning of the intense star formation activity triggered by the merging.

The evolution is divided in four phases: an intense star formation phase producing only high-mass stars (phase I), an early wind phase (phase II), a more quiescent phase producing stars in the whole mass range (phase III), and a last phase where star formation has stopped either because all the gas has been consumed or because a second wind occurred (Phase IV). The actual occurrence of an early wind phase and the possibility that it is followed by a new star formation phase will be discussed in section 5.1.

In phases I and III the number of stars created per mass (m) and time (t) intervals, $\Psi(t, m)$, normalized to the system mass $M_T$, is assumed to be of the form $\psi_i(t)\phi_i(m)$. $\psi_1(t)$ and $\psi_2(t)$ are the Star Formation Rates for the high-mass mode (Phase I) and normal mode (Phase III) respectively. The corresponding Initial Mass Functions, $\phi_i(m)$, are power laws of the stellar mass $\phi_i(m) \propto m^{-(1+x_i)}$ in the mass range $[m_l^i - m_u^i]$ and are normalized to 1. We will assume that the two modes IMF differ only by their low mass cut-off: for both modes we assume a classical Salpeter (1955) index $x_1 = x_2 = 1.35$ and a high-mass cut-off of $m_u^1 = m_u^2 = 100 M_\odot$.

In the first phase of the evolution we adopt a lower mass cut-off of $m_l^1 = 3 M_\odot$ (see the discussion in section 2) and a SFR decreasing exponentially with time, to restrict it, in principle, to the very beginning of the evolution:

$$\psi_1(t) = \nu_1 e^{(-t/\tau)} \quad with \quad \tau = 0.05 Gyr \tag{3}$$

However we will see that it is the occurrence of the outflow that actually limits the duration of this phase.

This phase ends when the thermal energy exceeds the binding energy, at time $t_{w1,b}$. Then an outflow starts (Phase II) that we parameterize as :

$$\frac{dM_{Gal}}{dt}(t) = -\alpha M_{Gal}(t) \tag{4}$$

$M_{Gal}$ is the mass of the galaxy at time t: $M_{Gal} = M_T - M_{Ej}$, where $M_{Ej}$ is the total mass ejected up to t. $\tau_{ML} = \alpha^{-1}$ is the characteristic time for the mass loss, which occurs at a decreasing exponential rate. During this period, we suppose that no star formation occurs: $\Psi(t, m) = 0$. This modeling is obviously very crude: it is likely that star formation has already decreased before the onset of the winds and on the contrary if the ISM is not homogeneous star formation could still continue in cold clouds embedded in the hot medium. A realistic model would require a full hydrodynamical modeling, where the wind would occur at different times as a function of the radius, coupled with the chemical evolution, on a multi-phase representation of the galaxy. A multi-phase model have been proposed by Ferrini and Poggianti (1993), at the expense of an increasing number of parameters. This is clearly beyond the scope of this paper.

When the thermal energy decreases below the binding energy, at time $t_{w1,f}$, the outflow stops and star formation begins again, producing stars in the whole mass range: $(m_l^2 = 0.1 M_\odot)$. In this phase III, we classically assume a SFR proportional to the gas mass available:

$$\psi_2(t) = \nu_2 \sigma_{gas}(t) \tag{5}$$

where $\sigma_{gas}(t) = M_{gas}(t)/M_T$.

Another wind may then happen, at time $t_{w2}$, due to the SN that will explode during this phase, when the gas content of the galaxy is low enough. The outflow that may happen afterwards will only involve a low gas mass. Therefore we will consider it as instantaneous and complete, as in the model of Matteucci and co-workers, and assume for simplicity that star formation will then cease (Phase IV), $\Psi(t, m) = 0$. We will see that in that phase the gas mass content is low and only a faint star formation could occur, which would not change our results.

Using the classical chemical evolution equations (Tinsley 1980; see also Prantzos et al. 1993), we follow the evolution of the mass contained in the gas (both in the ISM and in the outflow), of the mass contained in the form of evolving stars and remnants and compute the final luminosity of the galaxy (section 4.2). We also follow the metallicity in the gas and in the stars (mass averaged metallicity). In principle, this mass averaged stellar metallicity is not directly comparable to the observed metallicity, which is luminosity averaged. However it is a good approximation, as discussed in Matteucci (1993). The rates of type I and II supernovae (see section 4.3), and thus the thermal energy produced by the supernovae, as well as the binding energy of the gas are computed throughout the evolution (see section 4.4 and 4.5).

### 4.2. The Stellar Physics

The stellar lifetimes, $\tau(m)$, are taken from Schaller et al. (1992) for the solar metallicity (we shall see in section 5 that solar metallicities are reached rapidly).

We adopted the evolutionary tracks of stars computed in this same reference, for stars with $m > 0.8 M_\odot$, and used the main sequence luminosity given by Bruzual (1983) for lower mass stars. The bolometric luminosities and effective temperatures were converted to visual luminosity using the calibration of Böhm-Vitense (1981) for the main sequence stars and Flower (1977) for the Giants and Supergiants. We have neglected the effect of metallicity. Only stars born in phase III are still alive today ($t = 15 Gyr$). In addition, all stars more massive than $m_\tau$ have already died, where $m_t$ is the turn-off mass corresponding to $t - t_{w2}$, the time elapsed since the end of Phase III. The total final visual luminosity is then given by:

$$L_V = \int_{m_l^2}^{m_\tau(t-t_{w2})} \phi_2(m) dm \int_{t_{w1,f}}^{t_{w2}} \psi_2(t') L_V(m, t-t') dt' \quad (6)$$

where $L_V(m,t)$ is the visual luminosity of a star of mass $m$ at age $t$.

The masses of the stellar remnants, $R(m)$, are similar to those adopted by Prantzos et al. (1993): $R(m) = m$ for $m < 0.85 M_\odot$; $R(m) = 0.446 + 0.106 m$ for $0.85 < m/M_\odot < 9$, leading to white dwarf masses between 0.446 and $1.4 M_\odot$, according to the prescriptions of Iben and Tutukov (1984); $R(m) = 1.5 M_\odot$ for the neutron star remnants of stars with $9 < m/M_\odot < 25$; and $R(m) = 3 M_\odot$ for the black hole remnants beyond $25 M_\odot$.

The stellar yields are taken from Thielemann et al. (1992) for the stars more massive than $8 M_\odot$, and from Renzini and Voli (1981) for the low and intermediate mass stars. A quantitative idea of the heavy element production in each mode is obtained by integrating over the stellar yields, $y_n(m)$, the mass of element n produced by a star of mass m, weighted by the corresponding IMF:

$$Y = \int_{m_l^i}^{m_u^i} y_n(m) \, \phi_i(m) \, dm \quad (7)$$

Y is then the total mass of a heavy element which will be produced each time a unit of mass of gas is transformed into stars. All the stars born in the high-mass mode will have died by now, at variance with the normal mode. In that case, the lower bound of the integration was fixed at $1 M_\odot$, the present turn-off mass. The results are given in Table 1, where they are compared to the yields given by Woosley (1990), scaled to SN1987A for the iron production (0.07 $M_\odot$ for a $20 M_\odot$ star).

In spite of large differences between the models of Woosley (1990) and Thielemann et al. (1990), see the comparison made by Renzini et al. (1993), the integrated iron production is quite consistent, partly due to a common value for the $20 M_\odot$ star. This is not the case for other elements where discrepancies of about 40% are observed for Mg production and more than 50% for the O production.

**Table 1.** Integrated yields, weighted by the IMF

|  | $O^{16}$ | $Mg^{24}$ | $Fe^{56}$ |
|---|---|---|---|
|  | $10^{-3} M_\odot$ | $10^{-3} M_\odot$ | $10^{-3} M_\odot$ |
| Woosley (1990) | | | |
| high-mass mode | 70 | 2.8 | 2.6 |
| normal mode | 17 | 0.66 | 0.63 |
| Thielemann et al.(1990) | | | |
| high-mass mode | 44 | 2.0 | 2.4 |
| normal mode | 10 | 0.46 | 0.57 |

One notes that a $[Mg/Fe]$ value of 0.28 is expected from these estimates which do not take into account heavy element production by SNIa. The fact that this is very close to the value observed in ellipticals points towards a major role for SNII in the final stellar metallicity (the iron production by SNIa would reduce that value). Here and in the following we have assumed for reference an abundance by *mass* fraction in the solar neighborhood of $1.17 \times 10^{-3}$ for Iron and $5.8 \times 10^{-4}$ for Magnesium (meteoritic abundances, Anders & Grevesse 1989 and private communication).

Finally, the heavy element production by SNIa is also taken into account, adopting the yields obtained in the carbon deflagration model of Thielemann et al. (1986, model W7). Each SNIa produces 0.7 $M_\odot$ of Fe.

### 4.3. The SNIa and SNII rates

The rate of SNII is given by the number of dying stars more massive than $8 M_\odot$. It is null before $3.1\,Myr$, the lifetime of a $100\,M_\odot$ star and is given by the following equation afterwards:

$$R_{SNII}(t) = M_T \int_{sup(m_\tau(t), 8 M_\odot)}^{100 M_\odot} \Psi(t - \tau(m), m) dm \quad (8)$$

where $m_\tau(t)$ is the turn-off mass at time $t$.

In view of the uncertainties on the SNIa progenitors we adopt the convenient parameterization introduced by Ciotti et al. (1991) and assume that the first SNIa explodes after $0.1\,Gyr$:

$$R_{SNI}(t) = 1.1 \times 10^{-14} \, M_T \, r_{SNI} \, t_{15}^{-s} \quad yr^{-1} \quad (9)$$

where $t_{15}$ is the time in units of $15 Gyr$, the present time. We have normalized the SNIa rate to the initial galaxy mass, rather than to the final luminosity, which would have been more consistent: for a given $r_{SNIa}$ the final SNIa rate is not always the same in SNU units (1 SN/100 yr/ $10^{10}$ L$_{B\odot}$). $r_{SNI} = 1$ corresponds to the standard SNIa rate for ellipticals of 0.22 SNU (Tammann 1982) if we assume a final visual luminosity of 1.2 times the blue luminosity and a $M_T/L_V$ ratio of 17, which are the typical results we obtain (see Table 3). We shall generally assume that $s = 0$, following the argument presented in section 2, and $r_{SNI} = 0.25$, according to the new estimates of Cappelaro et al. (1993).

### 4.4. The binding energy of the ISM

Our computation of the binding energy of the gas, $E_{bgas}(t)$, follows the prescriptions of Matteucci (1992), based on the work of Bertin et al. (1992). If no dark halo is present and the luminosity profile follows the $R^{1/4}$ law (de Vaucouleurs profile):

$$E_{bgas}(t) = -\frac{1}{2}\frac{GM_L}{r_L}M_{gas} = \Omega M_{gas} \tag{10}$$

where $r_L$ is the half-mass radius, $M_L$ is the mass of the galaxy (star plus gas) and we have introduced for convenience the quantity $\Omega$, which is the equivalent of a gravitational potential. If $r_e$ is the effective (half light) radius of the galaxy then $r_L = 1.32 r_e$. Note that we have implicitly assumed here that the gas follows the stellar component although the gas distribution is likely to differ from the stellar one with time, since the gas absorbs the ejected supernova energy.

One further needs a relationship between the galaxy mass and $r_e$. Obviously, no observational constraint is known for a primeval galaxy. For present-day galaxies, the galaxy mass can be obtained from the virial theorem (Poveda 1958; Djorkovski and Davis 1987) and the observed central velocity dispersions $\sigma$ and effective radii (see section 3.2.2). From the data obtained by Djorkovski and Davis (1987) scaled to $H_0 = 50 km/s/Mpc$ and

$$M = 7.2 \times 10^2 \left(\frac{r_e}{pc}\right)\left(\frac{\sigma^2}{km/s}\right) \quad M_\odot \tag{11}$$

we derived the following correlation:

$$r_e = 17 \left(\frac{M_L}{10^{12} M_\odot}\right)^{0.585} \quad kpc \tag{12}$$

This is close to the relationship found by Saito (1979).

The $\Omega$ quantity thus scales as $M^{0.415}$. If this held also for the proto-galaxy and throughout the evolution its final value would be slightly smaller than its initial one: $\Omega_f = 0.75\Omega_i$, if the galaxy has typically lost half of it initial mass (see section 5). On the other hand, if the proto-galaxy has not yet fully collapsed, as is likely to be the case in Phase I, the final radius would be twice smaller than the initial radius, for a constant galaxy mass: $\Omega_f = 2 \Omega_i$ as discussed in Arimoto and Yoshii (1987,1989). The wind phase is also expected to change the effective radius versus mass relationship, but in the opposite direction. Under the assumption of homology and slow mass loss the product $Mr_e$ is a constant (Vader 1987), which would imply a final $\Omega_f = 1/4\Omega_i$. In view of all these uncertain factors we have not attempted to model the evolution of the gravitational potential and have kept it constant with time; it is then given by Eq.10 and the initial mass. The only time dependence of the binding energy is thus through the gas mass.

We have also considered the possible presence of a massive and diffuse dark halo. In that case the right term of Eq.10 is multiplied by a factor (Bertin et al. 1991) given as:

$$A = 1 + \frac{R}{\pi}\frac{r_L}{r_D}\left(1 + 1.37\frac{r_L}{r_D}\right) \tag{13}$$

where $R$ is the ratio of the dark matter mass to luminous mass and $r_L/r_D$ is the ratio of the radii of their respective distribution.

### 4.5. The thermal energy of the ISM

We only consider the supernovae contribution to the thermal energy of the gas and neglect the contribution of stellar winds. To compute the thermal energy we used the general formalism developed by Arimoto and Yoshii (1987) and Matteucci and Tornambé (1987). The thermal energy produced by a single SN depends on the local density of the ISM, via its cooling time. We thus explicitly take into account the gas density profile to compute the global energy, rather than adopting a mean value as previous authors.

The evolution with time of the thermal energy inside a SNR has been calculated by Cox (1972):

$$\epsilon(t) = \epsilon_0 \qquad \text{for } t < t_c$$
$$\epsilon(t) = \epsilon_0 \left[\frac{t}{t_c}\right]^{-0.62} \qquad \text{for } t > t_c$$

where $\epsilon_0 = 0.72 \times 10^{51}\ ergs$ and $t$ is the time elapsed since the SN explosion. The cooling time, $t_c$, depends on the local ISM density $n$:

$$t_c = 5.7 \times 10^4 n^{-9/17} \quad yr \tag{14}$$

where $n$ is expressed in cm$^{-3}$. Assuming that the gas follows a de Vaucouleurs density profile, as the stellar component, the density at the effective radius $r_e$ can be related to the total gas mass:

$$n(r_e) = 710 \left[\frac{M_{gas}(t)}{10^{12}\ M_\odot}\right]\left[\frac{r_e}{1\ kpc}\right]^{-3} \quad cm^{-3} \tag{15}$$

where we have cut the density profile at $15 r_e$. Adopting the relationship between effective radius and galaxy mass discussed in the previous section (Eq.12) we obtain:

$$t_c(r_e) = 1.59 \times 10^5 \left[\frac{M_T}{10^{12} M_\odot}\right]^{0.4} \sigma_{gas}^{-9/17} \quad yr \tag{16}$$

where $\sigma_{gas} = M_{gas}(t)/M_T$ is the total gas mass fraction. At other radii the cooling time scales as $t_c(r)/t_c(r_e) = [n(r)/n(r_e)]^{-9/17}$, a ratio which only depends on $r/r_e$.

The thermal energy per unit volume, $\epsilon_{th}(t, r)$, at radius $r$ and time $t$ is then:

$$\begin{aligned} \epsilon_{th}(t,r) &= \epsilon_0 \int_0^{t-t_c(r)} \rho_{SN}(t',r) \left(\frac{t-t'}{t_c(r)}\right)^{-0.62} dt' \\ &+ \epsilon_0 \int_{t-t_c(r)}^{t} \rho_{SN}(t',r) dt' \end{aligned} \quad (17)$$

where $\rho_{SN}(t,r)$ are the local SN rate per unit of volume. The star formation rate - hence the SN rate - depends explicitly on the local gas mass fraction for the low-mass mode ($\nu_2\,\sigma_{gas}$). This is not true in the case of the high-mass mode ($\nu_1\,exp(-t/\tau)$), but the latter can be considered as the expression of the integration over the volume of the galaxy of a function locally dependent on the ISM density. We thus assumed that the local rate is proportional to the local density: $\rho_{SN}(t,r)/\rho_{SN}(t,r_e) = n(r)/n(r_e)$ and $\rho_{SN}(t,r_e) = 3.54\,R_{SN}(t)/4\pi r_e^3$ for a de Vaucouleurs profile. Then Eq.17 must be integrated over volume to obtain the global thermal energy $E_{th}(t)$. A simple expression is obtained by replacing $t_c(r)$ by $t_c(r_e)$ in the integral boundaries of Eq.17. Most of the thermal energy comes from the contribution of SN in the already cooling phase (the cooling time is small as compared to other time scales so that $t \gg t_c$) and the exact value of $t_c$ in the definition of these boundaries is not important. We then obtained, from the assumed spatial variation of SN rate and cooling time:

$$\begin{aligned} E_{th}(t) &= 2.68\,\epsilon_0 \int_0^{t-t_c(r_e)} R_{SN}(t') \left(\frac{t-t'}{t_c(r_e)}\right)^{-0.62} dt' \\ &+ \epsilon_0 \int_{t-t_c(r_e)}^{t} R_{SN}(t') dt' \end{aligned} \quad (18)$$

Finally, during the wind phase, each time a given mass of gas is lost by the galaxy, the corresponding thermal energy is subtracted from the energy of the ISM.

## 5. Evolution of a typical elliptical galaxy

### 5.1. A Standard case

We discuss here the evolution of a typical elliptical galaxy of $5 \times 10^{11} M_\odot$. The parameters of the two IMFs were fixed to the values presented in section 4.1.: a common Salpeter slope and a lower cut-off at $3 M_\odot$ for the high-mass mode IMF. We suppose that there is no dark halo and a constant SNIa rate with $r_{SNIa} = 0.25$ (Eq.9).

The galaxy mass is close to the mass corresponding to the break luminosity in the Schechter luminosity function (section 7). Since the contribution to the total cluster luminosity of the galaxies located on one side of the break luminosity is equal to that of the other side, the iron mass ejected by such a galaxy, per unit of final luminosity, should give a good first approximation of the global contribution of all the galaxies to the ICM enrichment. The parameters $(\nu_1, \nu_2, \alpha)$ were thus chosen using the two main constraints: an ejected iron mass over final luminosity ratio which matches the ICM Iron mass over luminosity ratio found in clusters ($2 \times 10^{-2} M_\odot/L_\odot$), an iron abundance $[Fe/H]_\star$ in the galaxy stellar component of $\approx 0.15$ at the present time, observed for such a galaxy luminosity (see section 3.2). We thus adopt $(\nu_1, \nu_2, \alpha) = (40, 2, 18)\,Gyr^{-1}$.

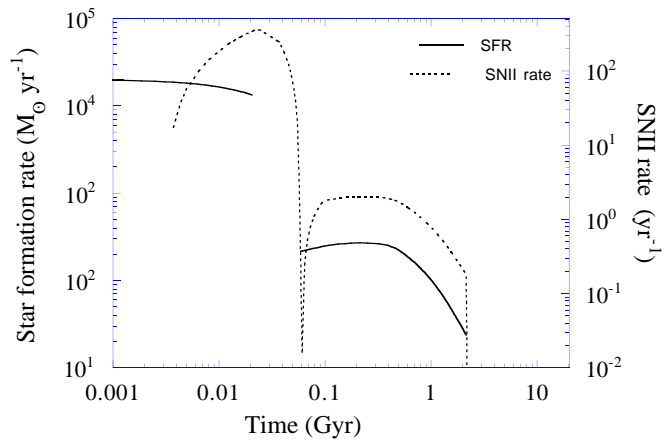

**Fig. 1.** Time variation of the SNII rate (dotted curve) and Star Formation Rate (full curve) for a galaxy of initial mass $M_T = 5 \times 10^{11} M_\odot$ and $(\nu_1, \nu_2, \alpha) = (40, 2, 18)\,Gyr^{-1}$.

The influence of the parameters on the final results will be discussed later, as well as their plausibility. For the clarity of these further discussions, we present here the general characteristics of the galaxy dynamical and chemical evolution in the model we consider, using this "standard case" as an example. The evolution of the main quantities are presented on Figures 1 to 5 and the main results are summarized in Table 3 (case C).

#### 5.1.1. The dynamical evolution and the origin of the different phases

The Star Formation and SNII rates are plotted on Fig.1, the binding and thermal energies on Fig.2. Fig.3 depicts the evolution of the galaxy mass, $M_{Gal}(t)$, the gas mass fraction inside the galaxy ($M_{gas}(t)/M_{Gal}(t)$) as well as the star and remnant mass fractions.

At the beginning of the evolution the gas is rapidly processed into high-mass stars with a SFR decreasing from 2 to $1 \times 10^4\,M_\odot/yr$. These stars begin to explode as SNII after $3.1\,Myr$, the lifetime of a $100\,M_\odot$ star. Since stars with the highest masses, which are less numerous, are the first to die, the SNII rate increases and would have kept increasing for $37\,Myr$, the lifetime $\tau_8$ of an $8\,M_\odot$ star, the lowest mass star dying as SNII. Before $\tau_8$ the thermal energy of the gas thus increases very rapidly following the SNII rate, while the binding energy of the gas, which is proportional to its mass, decreases. In other words, the specific thermal energy, defined as the ratio of the thermal energy to the gas mass, which has to be compared to the gravitational potential (Eq.10), keeps increasing due to a continuous and increasing input of energy by SNII in less and less gas. In that case the occurrence of a wind phase is automatic if $\tau_8$ is larger than the characteristic time $1/\nu_1$ of the star formation.

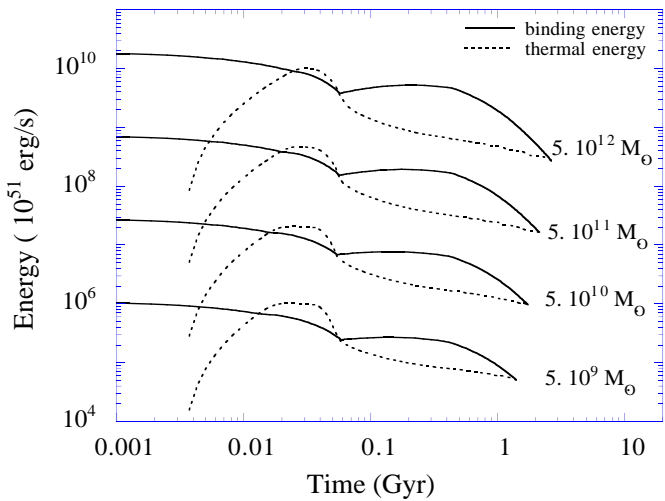

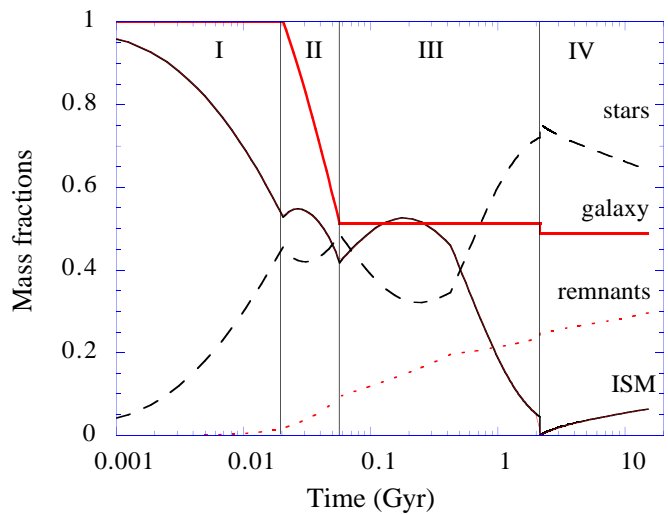

**Fig. 2.** Time variation with time of the gravitational binding energy of the gas (full line) as compared to its thermal energy (dotted line) for different galaxy masses.

**Fig. 3.** Variation with time of the mass fractions in the different components. Same galaxy mass and model parameters as in Figure 1. Heavy full line: galaxy mass as compared to the initial mass, $M_{Gal}(t)/M_T$. Full line: gas mass fraction inside the galaxy $M_{gas}(t)/M_{Gal}(t)$. Dashed line: living stellar mass fraction $M_*(t)/M_{Gal}(t)$. Dotted line: remant mass fraction $M_{rem}(t)/M_{Gal}(t)$. The thin vertical lines mark the different evolution phases : I = early burst phase; II = wind phase; III = quiescent star formation phase; IV = no star formation.

At $t = 20\ Myr = t_{w1,b}$, the thermal energy is higher than the binding energy and an outflow begins. At that time about 65% of the initial gas has been processed into stars, a fraction of which has already died and returned enriched gas to the ISM, and the galaxy contains by mass 53% of gas, 45% of stars and 2% of remnants.

During this outflow phase (Phase II), we assume that no star formation occurs. Therefore the evolution of the gas is ruled by two opposite phenomena: the death of stars, that feeds the ISM with processed gas, but also thermal energy from stars more massive than 8 $M_\odot$, and the outflow, which carries out of the galaxy both gas and thermal energy. The bump in the gas mass fraction, seen in Fig.3, is due to the gas restored by dying stars, which is larger than the mass loss of the galaxy at the beginning of the outflow phase. The SNII rate decreases steadily till $t = \tau_8$, the lifetime of an 8 $M_\odot$ star, because the progressive disappearance of the highest mass stars in the current Mass Function is somewhat compensated by the death of lower and lower mass stars still able to die as SNII. Afterwards the SNII rates decreases dramatically and would reach zero $\tau_8$ yrs after the onset of the wind phase and the turn off of star formation. Then the specific thermal energy begins to decrease significantly, due both to cooling and mass input from longer lived stars, which do not inject any thermal energy. It is thus clear here that the typical maximum duration of the wind in our model would be $\tau_8$ yrs. One must note that the wind duration inferred here is about 10 times less than the dynamical timescale of the galaxies and for a wind to be really established, the gas velocity should be highly supersonic. Highly supersonic early winds are also inferred in detailed hydrodynamical models of early type galaxies (David et al. 1990b, Ciotti et al. 1991). However it is clear that to examine further if our wind model is indeed quantatively self-consistent would require a full hydrodynamical treatment, with also a more detailed modeling of the gas cooling and heating. This is beyond the scope of this study and will be examined in a forthcoming paper.

The thermal energy of the gas is again lower than its binding energy at $t = 53\ Myrs$ and we suppose that the outflow stops (beginning of Phase III). At that time the galaxy has lost 49% of its initial mass and consists of 44% of gas, 47% of stars and 9% of remnants. The total mass lost in the wind clearly depends on the relative value of the mass loss time scale $(\tau_{ML} = 1/\alpha)$ as compared to the maximum duration of the wind $\tau_8$. In the case presented here they are of the same order. We can expect that for $\tau_{ML} \gg \tau_8$, the mass loss will be very low and for $\tau_{ML} \ll \tau_8$, all the gas restored by the stars more massive than $8M_\odot$ will be ejected. This will be discussed further in section 5.2.

At the beginning of the phase III, stars born in Phase I and less massive than $8M_\odot$ are still dying and feeding the galaxy with newly processed gas. It is thus reasonable to assume that star formation happens again, as we assume here. First, the ejection rate by the dying stars born during the high-mass phase is larger than the astration rate we adopted, provoking an increase of the gas mass (and hence of the SFR). The maximum SFR in that phase is 260 $M_\odot/yr$, much smaller than in Phase I. At $t = 0.44\,Gyrs$, the lifetime of a $3M_\odot$ star, the lowest mass star created in Phase I have died and the gas mass fraction decreases rapidly, on a time scale which depends on $\nu_2$. The SNII rate increases rapidly in a time $\tau_8$, as in Phase I, and then follows roughly the SFR. The evolution of the thermal energy in the gas is ruled by the re-heating due to these new SNII (plus SNIa) and global cooling and a slow decrease is observed. A second wind may appear when the gas mass fraction is low enough. This actually happens at $t = 2.1\,Gyrs$ when the gas only represents 5% of the galaxy mass.

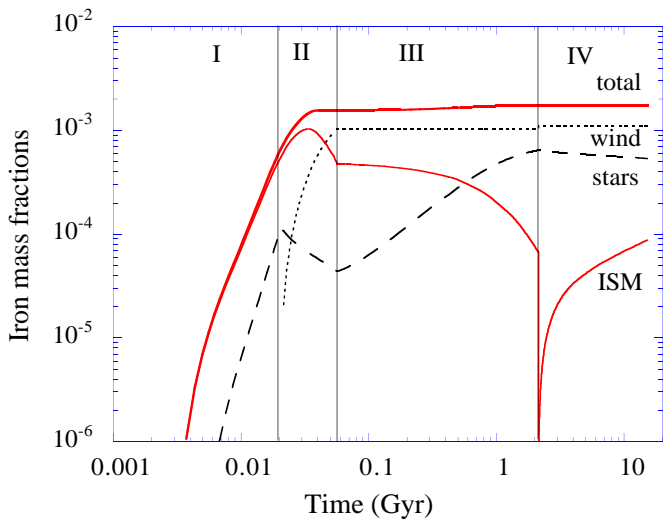
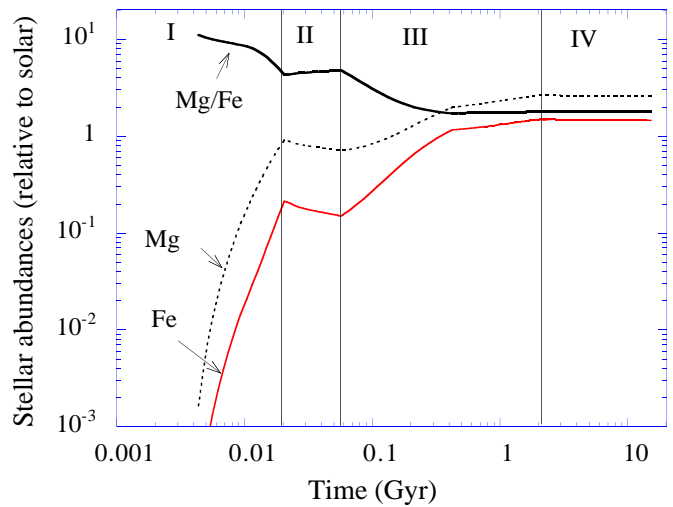

**Fig. 4.** Variation with time of the iron mass fractions. Same galaxy mass and model parameters as in Figure 1. Heavy full line: total iron mass fraction inside the system $M_{Fe}(t)/M_T$. Dotted line: iron mass fraction ejected outside the galaxy $M_{Fe}^{ej}(t)/M_T$. Dashed line: iron mass fraction in the stellar component $M_{Fe}^*(t)/M_T$. Full line: iron mass fraction in the Interstellar Medium $M_{Fe}^{ISM}(t)/M_T$

**Fig. 5.** Time variation of the Mg and Fe abundances relative to solar in the stellar component. Same galaxy mass and parameters as in Figure 1.

We simply assume then that all the gas mass that was left is ejected outside the galaxy and star formation stops definitively (Phase IV). The evolution of the galaxy is simply ruled by the secular death of stars: the gas and remnant mass fraction slowly increase, while the stellar mass fraction decreases.

At the present time, the galaxy has lost 51% of its initial mass and ends up with 6 % of gas, 64% of stars and 30% of dark matter in the form of stellar remnants with a $M/L$ ratio of $8.5 M_\odot/L_\odot$, perfectly consistent with the observed value (section 3.2.2). Its luminosity is $2.9 \times 10^{10} L_\odot$. The gas mass fraction is at the upper end of the values observed in ellipticals (section 3.2.2), however in our study we consider neither the possibility of another outflow, which may be driven by SNIa, nor further star formation in phase IV, which may proceed in the restored gas. Actually a faint star formation is still present in ellipticals today (Bregman et al. 1992, Fich & Hodge 1993; see also Nulsen et al. 1984 and Thomas et al. 1986, for considerations on the star formation in the cooling flow region of ellipticals).

### 5.1.2. The chemical evolution

We will first follow the distribution of the total iron mass produced by the galaxy and its share between the living stars, the ISM and the gas ejected out of the galaxy (Fig.4). These masses are normalized to the total initial mass of the galaxy $M_T$. Then we will discuss the time dependence of the Fe and Mg stellar (Fig.5) and ISM metallicities. We assume that the stellar ejecta mix instantaneously with the ISM.

The evolution of the total iron mass fraction is a direct consequence of the evolution of the SNII rates (Fig.1). It increases steeply till about $t = \tau_8$ and has already reached $1.5 \times 10^{-3}$ at the end of the wind phase when nearly all stars producing iron born in phase I have died. This quantity is directly the product of the fraction of gas which has been processed in Phase I multiplied by the iron production Y of the high-mass mode IMF as defined in section 4.2. The total iron mass fraction increases again in Phase II due to the enrichment by the normal mode and reaches $1.7 \times 10^{-3}$ at the end of the evolution. Thus 90% of the iron is produced by the massive stars born in the first phase of the evolution. The integrated contribution of the SNIa is negligible: $4 \times 10^{-5}$. Note also, that, in the framework of this model, any contribution from later stages of star formation possible star formation (i.e. in Phase IV, neglected here) would also be negligible.

In Phase I most of the iron is in the gaseous phase. The iron mass in stars increases steeply due to the astration of this enriched gas. During Phase II it decreases simply due the secular death of stars (no star formation). The total iron mass ejected from the galaxy increases following the outflow: the total iron mass ejected in the ICM at the end of the wind phase is $1. \times 10^{-3} M_T$, i.e. 3/4 of the iron mass produced then. The remaining iron is mostly in the galactic ISM. This iron, as well as the iron produced afterwards, is progressively decreasing while the one in the stellar component increases. At the end of phase III the stars contain $0.6 \times 10^{-3} M_T$ of Iron, the ISM $6.6 \times 10^{-5} M_T$. This last quantity is ejected in the ICM during the second wind. It is small as compared to the iron mass ejected during the first outflow. Some iron is restored to the ISM, as well as produced by SNIa, in Phase IV.

The galaxy ends with $0.5 \times 10^{-3} M_T$ of iron in stars, $0.09 \times 10^{-3} M_T$ in the ISM and $1.1 \times 10^{-3} M_T$ of iron has been ejected. From the luminosity obtained in the previous section the stellar iron mass to light ratio and ejected iron mass to light ratio are thus:

$$\frac{M^{ej}_{Fe}}{L_V} = 1.9 \times 10^{-2} \; \frac{M_\odot}{L_\odot} \quad and \quad \frac{M^*_{Fe}}{L_V} = 0.9 \times 10^{-2} \; \frac{M_\odot}{L_\odot} \quad (19)$$

These ratios crucially depend on the mass lost by the galaxy: if the total iron mass produced is ruled by the quantity of gas processed during the high-mass phase (and the Y quantity given by the IMF) its share between present stars and the ICM is determined by the mass loss, as well as the present-day luminosity (or mass) as compared to the initial mass. Therefore even if the present luminosity comes from a stellar population completely different from the stellar population which has created the iron ejected in the ICM, the total iron mass to light ratio, *together with* the ejected iron mass over stellar iron mass ratio, is a strong constraint on the model. Roughly speaking this last ratio constrains the proportion of mass lost by the galaxy, and thus its initial mass in view of the present luminosity. Then the total iron mass present tells us the global iron production from the stars born in the first burst phase.

The ISM metallicity increases till about $t = \tau_8$ following the SNII enrichment. The faster rise of the Mg abundance is a direct consequence of the yields we have chosen, which increase with star mass for Mg and are nearly constant for stars more massive than $20\ M_\odot$ for Fe (Thielemann et al. 1990). This behaviour is thus very model dependent: on the contrary Woosley (1990) predicts Fe yields increasing with stellar mass and the effect would have been less important. The ISM metallicity decrease till about 0.44 Gyr is due to a dilution effect: the injection of lower and lower metallicity gas by stars between $3\ M_\odot$ and $8\ M_\odot$, i.e. not producing Fe or Mg, created early in Phase I. At the beginning of Phase III this is not compensated by the enrichment from newly formed stars (a consequence of a returned mass input rate greater than the astration rate noted in the previous section). As a consequence the mean metallicity of the gas ejected during the wind phase is higher than the metallicity of the gas left for further star formation. This explains why, whereas the galaxy loses about 1/2 of its mass during the wind phase, it loses a larger proportion (about 3/4) of the iron mass available. This is not surprising in a model where the wind is driven by SNII, which produces both the required energy and the highly enriched gas. When all the stars created in Phase I have died, the ISM metallicity increases again and reaches 2.4 times the solar value (in number) at the present time. This is slightly higher than the maximum value allowed by the X-ray observations in nearby galaxies.

The mean stellar metallicity increases continuously in Phase I and III (continuous enrichment due to continuous star formation) and decreases in Phase II and to a less extent in Phase IV where the star formation is stopped and shorter lived stars which were created from higher metallicity gas die first. The final star metallicity is $[Fe/H]_\star = 0.15$, with a Mg over Fe ratio of $[Mg/Fe]_\star = 0.25$, consistent with the observational constraints given in section 3.2.1. This ratio is typical of an enrichment mainly by SNII (see the discussion in section 4.2) and is a direct consequence of our model.

### 5.2. Influence of the parameters

In this chapter, we restrict our investigation to the case of a typical galaxy mass of $5 \times 10^{11}\ M_\odot$ with no dark halo, while we will discuss the influence of the galaxy mass and presence of a dark halo in the next chapter. The parameters are given in Table 2 and the results summarized in Table 3 (case I-O).

#### 5.2.1. The early burst Star Formation Rate and mass loss rate

The major effect of a larger astration rate for the high-mass mode ($\nu_1$) is to increase the quantity of gas and heavy elements ejected by the galaxy: for $\nu_1 = 25\ Gyr^{-1}$ and $\nu_1 = 100\ Gyr^{-1}$ respectively (case I-J in Tables 2,3), 27% and 59% of the mass is lost in the wind; the iron mass ejected is $5.8 \times 10^{-4} M_T$ and $1.4 \times 10^{-3} M_T$. Obviously the final luminosity is smaller resulting in an even larger increase in the $M^{ej}_{Fe}/L_V$ ratio: $6.1 \times 10^{-3}$ and $3.3 \times 10^{-2}$ respectively. In parallel the $M^{ej}_{Fe}/M^*_{Fe}$ ratio increases from 0.23 to 1.9. One also notes a decrease of the mean stellar abundances and an increase of the fraction of remnants.

For higher $\nu_1$, the wind occurs earlier, because the specific energy of the gas increases more rapidly, but we found that about the same quantity of gas has been processed at the wind onset. A higher astration rate is compensated by a shorter phase. The global iron production is thus about constant. On the other hand, due to the intrinsic lifetime of massive stars, the returned mass fraction by stars has been smaller before the wind and the galaxy contains less gas at the onset of the wind. Then due to the larger concentration of SNII (less high-mass stars had time to die before the wind) on a smaller quantity of initial gas, the duration of the outflow is longer and a greater quantity of gas and heavy elements is ejected from the galaxy. The iron mass in the stellar component is smaller but not exactly in proportion to the final stellar mass: the final mean stellar abundances are smaller. This is a direct consequence of a smaller mean metallicity of the gas used for further star formation: due to a more intense wind phase where the richest star ejecta are expelled from the galaxy, the proportion in the ISM of gas restored by long lived stars created at low metallicity is larger (see previous section). The ratio $[Mg/Fe]_\star$ does not depend significantly on $\nu_1$: the enrichment is always driven by SNII.

**Table 2.** Model parameters

| Case | $M_G$ $M_\odot$ | $\nu_1$ $Gyr^{-1}$ | $\alpha$ $Gyr^{-1}$ | $\nu_2$ $Gyr^{-1}$ | $r_{SNI}$ SNU | $s$ | Dark Halo |
|---|---|---|---|---|---|---|---|
| A | $5\ 10^9$ | 40 | 18 | 2 | 0.25 | 0 | no |
| B | $5\ 10^{10}$ | 40 | 18 | 2 | 0.25 | 0 | no |
| C | $5\ 10^{11}$ | 40 | 18 | 2 | 0.25 | 0 | no |
| D | $5\ 10^{12}$ | 40 | 18 | 2 | 0.25 | 0 | no |
| E | $5\ 10^9$ | 40 | 18 | 2 | 0.25 | 0 | yes |
| F | $5\ 10^{10}$ | 40 | 18 | 2 | 0.25 | 0 | yes |
| G | $5\ 10^{11}$ | 40 | 18 | 2 | 0.25 | 0 | yes |
| H | $5\ 10^{12}$ | 40 | 18 | 2 | 0.25 | 0 | yes |
| I | $5\ 10^{11}$ | 25 | 18 | 2 | 0.25 | 0 | no |
| J | $5\ 10^{11}$ | 100 | 18 | 2 | 0.25 | 0 | no |
| K | $5\ 10^{11}$ | 40 | 8 | 2 | 0.25 | 0 | no |
| L | $5\ 10^{11}$ | 40 | 28 | 2 | 0.25 | 0 | no |
| M | $5\ 10^{11}$ | 40 | 18 | 1 | 0.25 | 0 | no |
| N | $5\ 10^{11}$ | 40 | 18 | 8 | 0.25 | 0 | no |
| O | $5\ 10^{11}$ | 40 | 18 | 2 | 1 | 1 | no |

The influence of $\alpha$ (case K and L, Tables 2,3), the coefficient of mass-loss of the outflow, which fixes the intensity of the wind is very similar. For higher $\alpha$, the mass lost per unit of time is larger, so that the SNII thermal energy is input into smaller quantities of gas and the specific energy is higher. The wind thus lasts even longer and due to this combination of higher mass loss for a longer time, more gas is ejected provoking a higher contribution to the ICM metal enrichment, a lower final luminosity, a lower stellar metallicity, and a higher ejected iron mass over stellar iron mass ratio.

In conclusion if the total iron mass produced per unit initial mass does not seem to be very dependent on the SFR, the ejected gas mass (and thus the final galaxy mass and luminosity) and the corresponding ejected iron mass are very sensitive to the astration rate and mass loss time scale. We find again clearly here that the global mass loss is the key parameter: it is tightly constrained to be $\approx 50\%$. It is clear that one can play with the 2 parameters $\nu_1$ and $\alpha$ (to decrease $\nu_1$ while increasing $\alpha$) so that this figure can be obtained. However there is a minimum value of $\nu_1$ of about $20\ Gyr^{-1}$ because in that case the early winds never occur ($1/\nu_1$ too large as compared to $\tau_8$, see section 5.1.1). Furthermore it is likely that the two quantities are physically related: for higher $\nu_1$ the SNII rate is larger during the outflow and it is likely that the mass loss rate is larger too. A strong point of our model is that once the mass loss is correct (as fixed for instance by the $M_{Fe}^{ej}/L_V$ ratio or equivalently by the $M_{Fe}^{ej}/M_{Fe}^\star$ ratio) the model predicts a correct value for the other fundamental quantity which is the total iron mass over luminosity ratio. This is because we have fundamentally a correct total iron mass production.

It is worthwhile to examine if the parameters favoured are realistic. A typical SFR of $\nu_1 = 40\ Gyr^{-1}$ is quite consistent with what is obtained for starburst galaxies: for M82 Doane & Mathews (1993) derived a star formation rate (assuming a truncated IMF at $3\ M_\odot$) $100-1000$ times that in the local neighbourhood, i.e. between $19\ Gyr^{-1}$ and $190\ Gyr^{-1}$, with best fit models corresponding to $\sim 50 Gyr^{-1}$. The mass loss rate $\dot{M}$ can be related to the SN rate, $R_{SN}$, through the energy equation (e.g. Vader 1987):

$$-\frac{1}{2}\dot{M}(v_e^2 + v_\infty^2) = \epsilon E_{SN} R_{SN} \qquad (20)$$

where $v_e^2 = 12\sigma^2$ is the escape velocity, directly linked to the gravitational potential, and $\epsilon < 1$ is the efficiency of conversion of energy released by each SN $E_{SN}$ into efficient energy for the gas escape. However the energy equation only provides an upper limit: the terminal wind velocity $v_\infty$ is unknown. For the high SN rates we obtained ($\sim 100-300\ SN.yr^{-1}$), we derived an upper limit for $\alpha = \dot{M}/M$ of $30-80\ Gyr^{-1}$, where we have used a $\sigma^2$ value deduced from Eq.11 and 12 and $\epsilon = 1$. So in the absence of a detailed hydrodynamical treatment, it is hopeless to constrain $\alpha$ further. All what can be said is that the typical value we considered, $\alpha = 18\ Gyr^{-1}$, has a correct order of magnitude, it is typically half the maximum value. We also checked that it is smaller than the critical value above which the mass loss of the galaxy would provoke the disruption of the galaxy, as given by Vader (1987). The expansion energy must be smaller than the kinetic energy so that $\dot{r}^2 < 3\sigma^2$, then assuming that $Mr$ is an invariant, one obtains for $\alpha = \dot{M}/M$:

$$\alpha < \alpha_c = \frac{\sqrt{3}\sigma}{r_e} \qquad (21)$$

**Table 3.** Model Results

| | Wind characteristics | | | | | | Galaxy present characteristics | | | | | | | |
|---|---|---|---|---|---|---|---|---|---|---|---|---|---|---|
| Case | $t_{w1}$ Myrs | $\Delta t_{w1}$ Myrs | $t_{w2}$ Gyrs | $f_{wind}$ | $M_{Fe}$ $M_\odot$ | $\left(\frac{O}{Fe}\right) \left(\frac{O}{Fe}\right)_\odot$ | $L_V$ $L_\odot$ | $\frac{M}{L_V}$ $\left(\frac{M}{L}\right)_\odot$ | $f_*$ % | $f_{rem}$ % | $f_{ISM}$ % | $\left[\frac{Fe}{H}\right]_*$ | $\left[\frac{Mg}{H}\right]_*$ | $\left(\frac{Fe}{H}\right)_{ISM} \left(\frac{Fe}{H}\right)_\odot$ |
| A | 14.0 | 45.0 | 1.40 | 0.61 | $4.33\ 10^6$ | 2.2 | $2.26\ 10^8$ | 8.7 | 65 | 27 | 8 | 0.06 | 0.29 | 2.0 |
| B | 16.1 | 38.6 | 1.75 | 0.54 | $4.50\ 10^7$ | 2.2 | $2.72\ 10^9$ | 8.5 | 65 | 27 | 7 | 0.12 | 0.35 | 2.2 |
| C | 20.5 | 36.5 | 2.14 | 0.51 | $5.49\ 10^8$ | 2.1 | $2.88\ 10^{10}$ | 8.5 | 64 | 30 | 6 | 0.16 | 0.41 | 2.4 |
| D | 25.0 | 32.5 | 2.67 | 0.46 | $6.08\ 10^9$ | 2.0 | $3.18\ 10^{11}$ | 8.4 | 64 | 31 | 5 | 0.21 | 0.47 | 2.6 |
| E | 18.3 | 36.5 | 1.96 | 0.52 | $4.91\ 10^6$ | 2.1 | $2.87\ 10^8$ | 8.4 | 65 | 28 | 7 | 0.15 | 0.39 | 2.3 |
| F | 22.8 | 34.5 | 2.40 | 0.49 | $5.81\ 10^7$ | 2.1 | $3.02\ 10^9$ | 8.5 | 64 | 30 | 6 | 0.19 | 0.44 | 2.5 |
| G | 28.4 | 28.5 | 2.98 | 0.42 | $6.19\ 10^8$ | 2.0 | $3.50\ 10^{10}$ | 8.3 | 64 | 31 | 5 | 0.25 | 0.51 | 2.7 |
| H | 35.1 | 15.2 | 3.38 | 0.25 | $4.39\ 10^9$ | 2.1 | $4.62\ 10^{11}$ | 8.1 | 66 | 30 | 4 | 0.37 | 0.61 | 3.2 |
| I | 41.8 | 14.5 | 2.20 | 0.27 | $2.88\ 10^8$ | 2.2 | $4.75\ 10^{10}$ | 7.7 | 68 | 25 | 7 | 0.26 | 0.51 | 2.5 |
| J | 7.79 | 46.5 | 2.27 | 0.59 | $7.23\ 10^8$ | 2.1 | $2.22\ 10^{10}$ | 9.3 | 60 | 35 | 5 | 0.06 | 0.26 | 2.3 |
| K | 20.5 | 28.1 | 2.19 | 0.24 | $2.56\ 10^8$ | 2.3 | $4.91\ 10^{10}$ | 7.8 | 70 | 23 | 7 | 0.24 | 0.49 | 2.4 |
| L | 20.5 | 40.1 | 2.38 | 0.69 | $7.54\ 10^8$ | 2.0 | $1.60\ 10^{10}$ | 9.7 | 54 | 41 | 5 | -0.10 | 0.22 | 2.3 |
| M | 20.5 | 36.5 | 5.58 | 0.50 | $5.34\ 10^8$ | 2.1 | $3.34\ 10^{10}$ | 7.5 | 67 | 30 | 3 | 0.19 | 0.43 | 2.8 |
| N | 20.5 | 36.5 | 0.45 | 0.55 | $5.70\ 10^8$ | 2.1 | $2.35\ 10^{10}$ | 9.6 | 58 | 31 | 11 | 0.16 | 0.40 | 2.0 |
| O | 20.5 | 36.5 | 2.04 | 0.51 | $5.81\ 10^8$ | 2.0 | $2.82\ 10^{10}$ | 8.6 | 64 | 30 | 6 | 0.30 | 0.41 | 8.5 |

Using Eq.11 and Eq.12 again, one can deduce the following value for $\alpha_c$:

$$\alpha_c = 30 \left[\frac{M}{10^{12} M_\odot}\right]^{-0.38} Gyr^{-1} \quad (22)$$

Hence, $\alpha$ must be lower than $40\ Gyr^{-1}$, for a $5 \times 10^{11}\ M_\odot$ galaxy.

### 5.2.2. The quiescent Star Formation Rate

The final results are largely insensitive to the low-mass mode astration rate $\nu_2$, as can be seen in Tables 2,3 (case M-N) where we have let vary $\nu_2$ in a wide range: between $\nu_2 = 1\ Gyr^{-1}$ and $\nu_2 = 8\ Gyr^{-1}$. By increasing $\nu_2$, a smaller stellar metallicity is obtained, due to a shorter phase III (the second wind occurs earlier) and thus less further enrichment by the normal IMF stars. The effect is small however because the enrichment process is dominated by the high-mass mode, as seen above. More gas is ejected during the second wind and the total iron mass ejected in the ICM is slightly larger. The final $M/L$ is 20% larger because the present stellar population is older. The effect on the $M^{ej}_{Fe}/L_V$ ratio is small: $1.6 \times 10^{-2}$ and $2.4 \times 10^{-2}$ for respectively $\nu_2 = 1\ Gyr^{-1}$ and $\nu_2 = 8\ Gyr^{-1}$. These values, as well as the mean stellar metallicity obtained are well within the range allowed by the observations. It is thus clear that this is not a crucial parameter of our model.

### 5.2.3. The SNIa Rate

If a high SNIa rate in the past ($s = 1$, Eq.9) together with a standard present SNIa rate of 0.22 SNU are assumed, the SNIa play a significant role in the iron enrichment in Phase III (see Tables 2,3; case O). A significant higher stellar iron abundance is obtained with $[Fe/H] = 0.3$ and naturally a lower ratio $[Mg/Fe] = 0.1$, closer to the solar value. Both values are hardly compatible with the observational constraints (see section 3.2.1). In addition, as expected from the discussion in section 2, the predicted iron abundance in the present ISM is much too large: 8 times the solar value. Thus in the framework of a bimodal star formation model, the past SNIa rate of ellipticals must be close to the present one.

## 6. Evolution versus galaxy mass

We present here the influence of the galaxy mass on the results, for the standard set of parameters defined in section 5.1.

Due to a deeper gravitational potential, the early wind as well as the second wind occur later for more massive galaxies (see Fig.2). The global iron production thus increases with the galaxy mass: from $1.3 \times 10^{-3} M_T$ to $2.0 \times 10^{-3} M_T$ for a $5 \times 10^9\ M_\odot$ galaxy and a $5 \times 10^{12}\ M_\odot$ galaxy respectively. For the same reason the duration of the outflow is smaller and more massive galaxies lose less mass: 61 % for a $5 \times 10^9 \, M_\odot$ as compared to 46 % for $5 \times 10^{12} \, M_\odot$ galaxy. The ejected iron mass fraction is slightly larger, however, due to the higher SNII rate induced by a longer high-mass star formation phase and the final $M_{Fe}^{ej}/L_V$ ratio is nearly independent of the galaxy mass. The $M/L$ ratio is nearly constant, it even slightly decreases with the galaxy mass: the slightly larger proportion of remnants due to a longer high-mass phase is compensated by a higher luminosity due to a younger stellar population.

Due to a higher metal production the final stellar metallicities increase with the galaxy mass: from 1.1 times solar to 1.6 times solar for Fe; from 1.9 time solar to 2.9 time solar for Mg, but the $[Mg/Fe]_\star$ does not vary significantly. This is again a consequence of an enrichment dominated by SNII.

Thus if the model does predict an increase of metallicity with galaxy mass (or luminosity) it is clearly unable to account for a significant increase of the $M/L$ and $[Mg/Fe]_\star$ ratio. As discussed in section 3.2 the general increase of these ratios with mass is uncertain, however, should they be confirmed, our model as it stands would be rejected.

As discussed in Renzini and Ciotti (1993), reproducing the trend of $M/L$ ratio with $L$, while keeping a relatively small and constant dispersion at each location of the Fundamental Plane, is not straigthforward. It may require systematic variations of either the IMF or the dark matter content of galaxies, i.e systematic variations of either $M_\star/L$ or $M/M_\star$, where $M_\star$ is the mass of the bright component. Possible variations of the IMF are not taken into account in our standard model. One possibility would be to increase the proportion of high-mass stars in the high-mass mode IMF for more massive galaxies (by varying either the slope or the low mass cut-off). However we found that the effect is small on both the $M_\star/L$ and the $[Mg/Fe]_\star$ ratio. A variation of the dark matter content (of the parameters $R$ and $r_L/r_D$ defined section 4.4 and considered by Renzini and Ciotti 1993) is a more likely explanation: in the framework of our model, we checked that this would not change dramatically the final results, in terms of ejected iron mass over luminosity ratio and stellar metallicity as well as $M_\star/L$ ratio (see below and Tables 2,3 cases E-H).

On the other hand the possible increase of the $[Mg/Fe]_\star$ ratio with mass is more difficult to explain in the framework of our model. The only way we see to account for it, is to increase the role of SNIa (see case O). It would be larger for more massive galaxies due to a longer normal phase. However this sets two problems: first the ISM metallicities would be too large (see previous section) and it would predict a larger variation of the Fe stellar abundance than that of the Mg abundance, whereas the observations seem to indicate the contrary (see section 3.2.1).

Finally we considered the possibility that the galaxies may be embedded in a dark halo (Tables 2,3, case E-H, where the quantities $M_G$ and $M/L$ refer to the baryonic component only). We assumed 10 times more mass in the dark halo than in the luminous component, with a typical radius of the dark matter 5 times that of the luminous component ($R = 10$ and $r_L/r_D = 0.2$ in Eq.13). The net effect is the same as obtained by increasing the galaxy mass which is not surprising: higher stellar metallicities and ejected iron mass fraction. Using the same standard set of parameters defined in section 5.1, somewhat too high stellar metallicities are obtained, but the increase is not drastic (typically 0.1 dex). The same small effect was noticed by Matteucci (1992) in the framework of her model. The final $M_{Fe}^{ej}/L_V$ ratio are 20% smaller than in the case of no dark halo. More correct results can be easily obtained by changing slightly the mass loss parameter.

In the following discussion we will adopt the models A to D, which proved to account essentially for the observational constraints on elliptical galaxies we discussed in section 3.2.

## 7. The intra-cluster medium

To compute the total contribution of the winds to the ICM, we integrate the contribution of each galaxy, assuming that the luminosity distribution of the galaxies in a cluster follows the Schechter luminosity function (Schechter 1976):

$$N(L)dL = N^* \exp(-L_V/L_V^*)L^{-\alpha}dL \qquad (23)$$

where we adopt $\alpha = 1.25$ and a break luminosity $L_V^* = 4.5 \times 10^{10} L_\odot$. This integration is trivial, as shown by Arnaud et al. (1992) if the quantity considered (e.g. ejected gas mass) is a power law of the (present) visible luminosity of the galaxy:

$$M_X = M_X^* \left[\frac{L_V}{L_V^*}\right]^a \qquad (24)$$

where $M_X^*$ is the value obtained for a galaxy of luminosity $L_V^*$. In that case the ratio of this quantity integrated over the cluster population to the total cluster luminosity, $L_V$, is:

$$\frac{M_X}{L_v} = \frac{M_X^*}{L_V^*} \frac{\Gamma(-\alpha + a + 1)}{\Gamma(-\alpha + 2)} \qquad (25)$$

Such a power law correctly fits the results obtained for different initial galaxy masses (section 6 and Tables 2,3; cases A-D), as shown on Fig.6 where the ejected gas and iron mass are plotted versus the final luminosity of the galaxy. We obtained an ejected gas mass of 8.4 $L_V^*$ $(M_\odot/L_\odot)$ for the galaxy corresponding to the break and a power index of $a = 0.92$; the corresponding values for the ejected iron mass are $1.86 \times 10^{-2} L_V^*$ $(M_\odot/L_\odot)$ and $a = 1.00$.

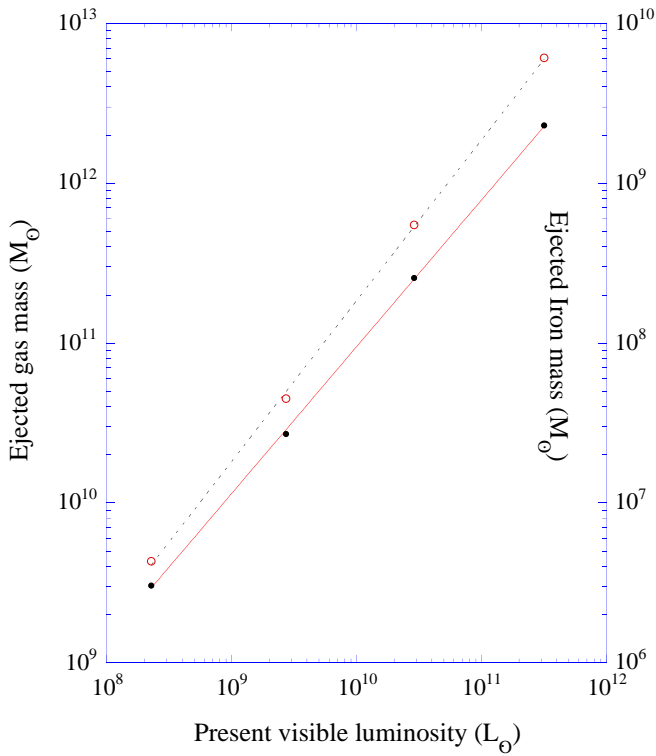

**Fig. 6.** Gas mass (full line) and iron mass (dotted line) ejected in the intra cluster medium versus the present visible luminosity of the galaxy. The four points correspond to the model case A to D (see Tables 2,3). The line is the best fit to the results.

The final ratio of ejected gas mass over total cluster luminosity is then:

$$\frac{M_{gas}^{ej}}{L_V} = 9.3 \, \frac{M_\odot}{L_\odot} \quad (26)$$

and the ratio of ICM iron mass to luminosity is:

$$\frac{M_{Fe}^{ICM}}{L_V} = 2 \times 10^{-2} \frac{M_\odot}{L_\odot} \quad (27)$$

The contribution of all the cluster galaxies to the ICM enrichment is perfectly consistent with the observational data (Eq.1), as was expected from the value of iron mass ejected by the typical elliptical studied in section 5. This is not surprising since the ejected iron mass appears to be proportional to the luminosity ($a = 1.00$).

Although our model predicts more mass loss than others, we confirm that most of the gas must be of primordial origin, as already found by previous authors ( Matteucci & Vettolani 1988; David et al. 1991b; Arnaud et al. 1992). From Eq.26 and Eq.2 we deduce a mean proportion of primordial gas in the ICM of 70%, ranging from $\approx 50$ % for groups of galaxies to $\approx 80$ % for rich clusters. As a consequence, one can easily understand the observed decrease of the iron abundance with the cluster temperature (or richness), first shown by Hatsukade (1989), as a consequence of the larger dilution of the processed gas ejected by the cluster galaxies in rich clusters. This dilution effect was already found in previous models (David et al. 1991b; Arnaud et al. 1992).

Finally as expected in a model based on an enrichment by SNII, we obtain an $\alpha$ element to iron ratio above solar in the ICM:

(O/Fe)   =   $2.1 \, (O/Fe)_\odot$
(Si/Fe)  =   $1.3 \, (Si/Fe)_\odot$
(Mg/Fe)  =   $1.8 \, (Mg/Fe)_\odot$

This is in agreement with the observations of Canizares et al. (1982, 1988) but these observations should still be confirmed. Unfortunately this requires the temperature to be low enough for the elements below Fe not to be completely ionized; this is the case in groups or in the cooling flows present at the center of some clusters. ASCA has the sensitivity and spectral resolution necessary to perform such measurements, and will provide a crucial test on the respective roles of SNIa and SNII in the enrichment of the ICM.

## 8. Conclusion

We showed that a bimodal star formation model, where only high-mass stars are created in a violent star formation phase in the initial stage of evolution, can be applied to elliptical galaxies, considered as responsible for the iron enrichment of the ICM. The idea of a truncated IMF is supported by observations of starburst galaxies, where the star formation activity seems to be linked with the proportion of high-mass stars. The model favour SNII as the major contributors for both the Fe production and heating of the ISM of the underlying galaxies, at the expense of SNIa. Most of the iron production is due to the massive stars created in the early burst phase. The advantage of such a model is to provide a fast and large concentration both in space and time of SNII, which induces a large outflow of enriched gas towards the ICM. Part of the iron is ejected in the ICM during this wind phase and the remaining iron is locked afterwards in the later generations of star created with a normal IMF.

The ICM iron mass per unit luminosity of member galaxies ($2 \times 10^{-2} \, M_\odot/L_\odot$) is reproduced, together with other observed quantities of present elliptical galaxies ($M/L$, $[Mg/Fe]_*$ higher than solar, stellar and ISM metallicities, remaining gas fraction) for reasonable values of the parameters (astration rate in the burst phase and in the quiescent phase, mass loss rate during the wind phase).

Even if the present luminosity comes from a stellar population completely different from the stellar population which has created the iron ejected in the ICM, we showed that the total iron mass to light ratio in clusters, *together with* the ICM iron mass over stellar iron mass ratio, is a strong constraint. The share of the iron produced between present stars and the ICM is determined by the mass loss, as well as the present-day luminosity (or mass) as compared to the initial mass. A key factor is thus the total mass lost during the wind. It must be $\approx 50\%$ in order to insure the observed even share between present stars and the ICM, as determined by the observed $M_{Fe}^{ej}/M_{Fe}^{*}$ ratio. A strong point of the model is that once the mass loss is consistent with this observation, a correct value for the other fundamental quantity which is the total iron mass over luminosity ratio is obtained. This is because the model fundamentally gives a correct total iron mass production per unit initial mass of the galaxy. We found that this quantity is very robust with respect to the astration rates: it is determined by the quantity of gas processed during the high mass phase, which is regulated by the occurrence of the wind, multiplied by the global yield of the truncated IMF. This further supports our choice of this type of IMF.

The evolution depends on the galaxy mass due to the deeper gravitational potential for larger galaxies. A possible weak point of the model is that, although metallicities increase with galaxy mass, other quantities ($Mg/Fe$ and $M/L$) vary only slightly.

It is concluded, as in previous works, that most of the cluster gas ($\approx 80\%$, for rich clusters) is of primordial origin, which explains the decrease of iron abundance with cluster temperature, itself related to the richness of the cluster. A decisive test for the model will be given by the observational determination of the abundances ratios in clusters of Si, O or Mg over Fe that we predict to be larger than solar by respectively a factor of 1.3, 1.7 and 2.1. Such observations, possible in poor clusters or in cooling flows with X-ray missions like ASCA or XMM, will be determinant in the next years.

*Acknowledgements.* We are very grateful to M.Cassé and N.Prantzos for very fruitful discussions and careful reading of the manuscript. N.Prantzos kindly provided us with his numerical code for galaxy chemical evolution. We would also like to thank W. Forman for his useful comments on the manuscript and A. Renzini for interesting discussions on the relative role of SNI and SNII and for his referee report, which helped us to improve this paper.